\documentclass[twocolumn,twocolappendix,tighten]{aastex63}

\received{}
\revised{}
\accepted{}

\usepackage{amssymb}
\usepackage{amsmath}

\def \half {\textstyle{\frac{1}{2}}}

\def \p {\partial}
\def \d {\partial}

\def \. {\cdot}
\def \x {\times}

\newcommand{\kms}{km~s$^{-1}\,$}
\newcommand{\alfven}{Alfv{\'e}n\,}

\begin{document}
\title{A Cancellation Nanoflare Model for Solar Chromospheric and Coronal Heating III. 3D Simulations and Atmospheric Response}
\correspondingauthor{P. Syntelis}
\email{ps84@st-andrews.ac.uk}

\author{P. Syntelis}
\affiliation{St Andrews University, Mathematics Institute, St Andrews KY16 9SS, UK}

\author{E.R. Priest}
\affiliation{St Andrews University, Mathematics Institute, St Andrews KY16 9SS, UK}

\begin{abstract} 

Inspired by recent observations suggesting that photospheric magnetic flux cancellation occurs much more frequently than previously thought, we analytically estimated the energy released from reconnection driven by photospheric flux cancellation, and proposed that it can act as a mechanism for chromospheric and coronal heating \citep{Priest_etal2018}. 
Using two-dimensional simulations we validated the analytical estimates and studied the resulting atmospheric response \citep{Syntelis_etal2019}. 
In the present work, we set up three-dimensional resistive MHD simulations of two cancelling polarities in a stratified atmosphere with a horizontal external field to further validate and improve upon the analytical estimates.
The computational evaluation of the parameters associated with the energy release are in good qualitative agreement with the analytical estimates. 
The computational Poynting energy flux into the current sheet is in good qualitative agreement with the analytical estimates, after correcting the analytical expression to better account for the horizontal extent of the current sheet.
The atmospheric response to the cancellation is the formation of hot ejections, cool ejections, or a combination of both hot and cool ejections, which can appear with a time difference and/or be spatially offset, depending on the properties of the cancelling region and the resulting height of the reconnection. Therefore, during the cancellation, a wide spectrum of ejections can be formed, which can account for the variety of multi-thermal ejections associated with Ellerman bombs, UV bursts and IRIS bombs, and also other ejections associated with small-scale cancelling regions and spicules.

\end{abstract}


\section{Introduction} 
\label{sec:introduction}

Coronal heating is one of the biggest open questions in solar and stellar physics. Many ideas have been proposed on how the solar corona is being heated to multi-million degrees and the chromosphere to tens of thousands of degrees, with most models considering that the energy required  originates either from waves \citep[e.g., resonant absorption or phase mixing][]{Klimchuk_etal2006,Parnell_DeMoortel_2012,Priest_2014} or from magnetic reconnection \citep[e.g., nanoflares driven by photospheric motions][]{Parker_etal1988,Priest_etal2002}.
Yet, no mechanism has conclusively been identified as the one heating the solar corona.

In recent years, observations of the solar photosphere and atmosphere have revealed that the cancellation of opposite magnetic polarities is a much more common process than previously thought. Magnetic flux cancellation is the process whereby two opposite polarities approach each other, interact via magnetic reconnection and eventually submerge into the solar interior \citep[e.g.][]{Parker_1979, vanBallegooijen_etal1989}. The reconnection associated with flux cancellation has been long proposed as a mechanism for heating X-ray bright points \citep[e.g.][]{priest94b,parnell95}, but the great increase in frequency of cancellation events suggests that they may also be heating chromospheric and coronal plasma much more widely. 

Observations of the plasma properties above cancellation regions have revealed that both cool and hot jets can be ejected above the cancelling polarities as a result of the reconnection driven by the cancellation. Depending on the height where that reconnection occurs, the localised energy release and resulting plasma ejections may show up as Ellerman bombs, UV and EUV bursts, IRIS bombs and other impulsive releases of energy \citep[e.g.][]{
Watanabe_etal2011,
Vissers_etal2013,
Peter_etal2014,
Kim_etal2015,Vissers_etal2015,Rutten_etal2015,Rezaei_2015,
Nelson_etal2016,Tian_etal2016,Reid_etal2016,Rutten_2016,
Nelson_etal2017,Toriumi_etal2017,Hong_etal2017,Libbrecht_etal2017,RouppevanderVoort2017,Tian_etal2018, Ulyanov_2019, Huang_etal2019, Chen_etal2019}.
In addition to the above energy bursts and plasma ejections, recent observations have suggested that the cancellation of opposite polarities at the footpoints of coronal loops could be responsible for the brightening of coronal loops \citep[][]{tiwari14,chitta17b,huang18,chitta18,Sahin_etal2019}.

State of the art observations from the Sunrise balloon mission \citep{solanki10a,solanki17b} measured the photospheric magnetic field with a spatial resolution of 0.15 arcsec, six times the resolution of the Helioseismic Imager (HMI) on the Solar Dynamics Observatory (SDO). 
These novel observations revealed that the rate of magnetic flux cancellation is higher than previously thought by an order of magnitude \citep{smitha17}. These findings suggest that the energy released during these previously unseen cancelling flux elements is more ubiquitous than previously realised.

Inspired by these observations, as a first step towards estimating the energy release, we analytically examined  the cancellation of two opposite polarities \citep[][hereafter Paper I]{Priest_etal2018}. 
By assuming that the two converging opposite polarities are in the presence of a uniform horizontal magnetic field, we estimated the energy released during the reconnection driven by the cancellation, and the height of release.
Our analysis led us to propose that reconnection driven by the photospheric flux cancellation is a nanoflare-based mechanism able to heat the solar chromosphere and corona.

The next step in  developing the model was to set up two-dimensional (2D) simulations of flux cancellation that have the same features as our analytical model \citep[][hereafter Paper II]{Syntelis_etal2019}, so that they may be compared. By assuming two converging opposite magnetic polarities inside an overlying uniform horizontal magnetic field, we compared the analytical expressions with the simulations. We found that our analytical expressions accurately estimated the characteristic parameters associated with the reconnection region, such as the magnitude of the inflowing magnetic field and velocity,  the length of the current sheet, and the resulting energy converted to heat. 
In addition, because the simulations included a stratified atmosphere, we were able to study the atmospheric response to the reconnection occurring at different heights. We found that depending  on  the initial height of the null point, the cancellation-driven reconnection could produce either hot or cool ejections or a combination of both hot and cool ejections, in a manner similar to observations.

In this paper, we  take the next step towards developing our model by setting up three-dimensional (3D) simulations of reconnection driven by flux cancellation. To do so, we consider two converging opposite polarities in the presence of an overlying horizontal magnetic field; a field similar to that of Paper I. We again include a stratified atmosphere, to compare with the 2D simulations.

The paper is  organised as follows. 
In Section \ref{sec:theory} we summarise the analytical theory presented in Paper I. 
In Section \ref{sec:numerical_simulations} we describe our computational model, compare it with the analytical theory, and examine the atmospheric response. 
In Section \ref{sec:discussion} we discuss our results.

\section{Summary of Theory for Energy Release driven by Photospheric Flux Cancellation in 3D} 
\label{sec:theory}

Here we summarise the theoretical estimates of the energy release by steady-state magnetic reconnection driven by flux cancellation in three dimensions, presented in Paper I.

\subsection{Energy Conversion during Photospheric Flux Cancellation in a Horizontal Field} 

\subsubsection{Magnetic configuration}

At the photosphere, we considered two sources, one with positive and one with negative magnetic flux ($\pm F$), situated at points  B $(d,0,0)$ and A $(-d,0,0)$ on the $x$-axis.  In the atmosphere, we included a uniform and horizontal magnetic field $B_{0}\bf{\hat{x}}$. 
For simplicity, we assumed that the sources have equal flux and that they were aligned with the overlying field (see Paper I for discussion on more general configurations).
The resulting magnetic field is given by:

\begin{equation}
{\bf B}=\frac{F}{2\pi}\frac{{\bf \hat{r}_{1}}}{r_{1}^2}-\frac{F}{2\pi}\frac{\bf{\hat{r}_{2}}}{r_{2}^2}+B_{0}{\bf{\hat{x}}},
\label{eqn23}
\end{equation}
where
\begin{equation}
{\bf r}_{1}=(x-d){\bf \hat{x}}+y{\bf \hat{y}} + z{\bf \hat{z}}, \ \ \ \ \ \ \ 
{\bf r}_{2}=(x+d){\bf \hat{x}}+y{\bf\hat{y}} +z{\bf \hat{z}}, \nonumber
\label{eqn24}
\end{equation}
are the vector distances from the two sources to a point $P(x,y,z)$.

The magnetic field is non-dimensionalised with respect to $B_0$, and the length with respect to the interaction distance \citep{Longcope_1998}
\begin{equation}
    d_0 = \left( \frac{F}{\pi B_0} \right)^{1/2},
\end{equation}
and so we define dimensionless quantities:
\begin{align}
{\bar{B_{x}}}=\frac{B_{x}}{B_{0}},\ \  
{\bar{d}} &=\frac{d}{d_{0}},\ \  
{\bar{\bm r}}=\frac{\bm r}{d_{0}},\ \ \\
{\bar{x}}=\frac{x}{d_{0}},\ \ 
{\bar{y}}&=\frac{y}{d_{0}},\ \ 
{\bar{z}}=\frac{z}{d_{0}}. 
\nonumber
\end{align}

The $x$-component of the field then becomes:
\begin{equation}
 \bar{B}_{x} = 
 \frac{\bar{x}-\bar{d}}{2\bar{r}_{1}^{3}}-\frac{\bar{x}+\bar{d}}{2\bar{r}_{2}^{3}}+1,
 \label{eqn9}
\end{equation}

Along the $z$-axis therefore, at $\bar{x}=\bar{y}=0$, the $y$- and $z$- components of the magnetic field vanish ($\bar{B}_{y}=\bar{B}_{z}=0$), while the $x$-component becomes
\begin{equation}
 \bar{B}_{x} = 
 -\frac{\bar{d}}{(\bar{d}^{2}+\bar{z}^{2})^{3/2}}+1.
 \label{eqn10}
\end{equation}

\begin{figure*}
\begin{center}
\includegraphics[width=\linewidth]{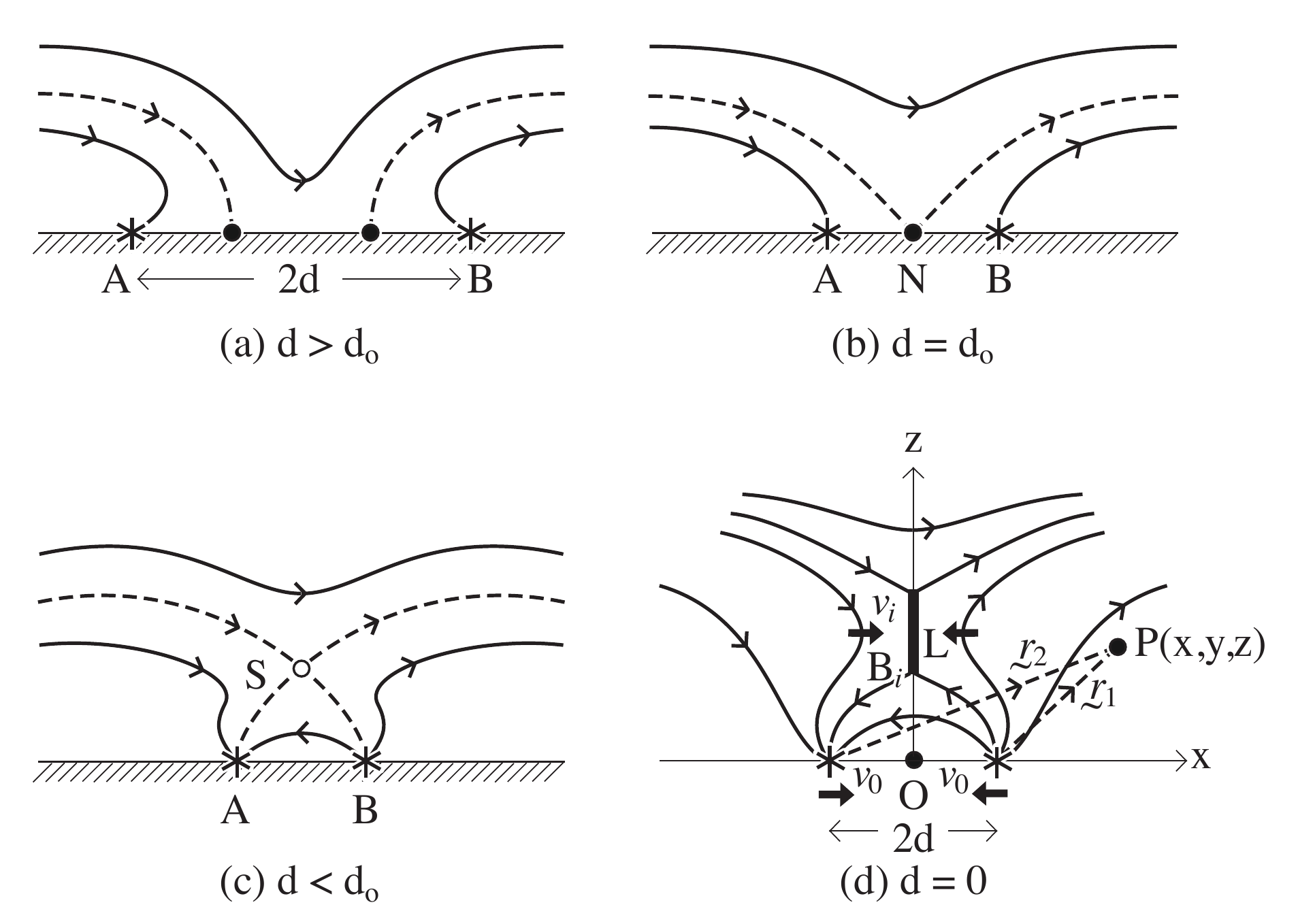}
\caption{
Magnetic topology in the $xz$-plane during reconnection driven by photospheric flux cancellation when (a) $\bar{d}\equiv d/d_0>1$, (b) $\bar{d}=1$ and (c) $\bar{d}<1$. (d) Schematic showing the notation used to describe the reconnection region. Figure taken from Paper I.
}
\label{fig:cartoon}
\end{center}
\end{figure*}

Suppose the two sources  approach one another at 
speeds $\pm v_{0}$. 
The evolution of the magnetic field topology is described in detail in Paper I, and is schematically represented in Figure~\ref{fig:cartoon}. 
The two sources are not magnetically linked when they are far away ($\bar{d} >1$, Figure~\ref{fig:cartoon}a).
When $\bar{d}=1$, a high-order null point appears at the origin (point N, Figure~\ref{fig:cartoon}b). As the sources approach each other
($\bar{d}<1$), a semi-circular separator is formed in the $yz$-plane at $x=0$ (marked as S, Figure~\ref{fig:cartoon}c). 
The radius ($\bar{z}_{S}$) of the  separator  is found from Equation (\ref{eqn10}) by setting $\bar{B}_x = 0$, namely,
\begin{equation}
\bar{z}_{S}^{2}=\bar{d}^{2/3}-\bar{d}^{2},
\label{eqn11}
\end{equation}
and ${z}_{S}$ is plotted in Figure 4a of Paper I. As the sources approach, the radius of the separator increases to a maximum of 
$(\bar{z}_s)_{max} = (4/27)^{1/4}$ at $\bar{d}=1/3^{3/4}$ and then drops to zero as $\bar{d}$ tends to zero.

\subsubsection{Inflow Plasma Speed ($v_i$) and Magnetic Field ($B_i$) at the Reconnection region}

During the reconnection driven by flux cancellation, a semi-annular current sheet is formed at the location of the separator. In the $xz$-plane, this corresponds to a sheet of length $L$ (Figure~\ref{fig:cartoon}d). 
By linearising the field around the current sheet, the magnetic field strength of the plasma flowing into the current sheet is found to be 
\begin{equation}
\bar{B}_{i}=\frac{3 (1-\bar{d}^{4/3})^{1/2}}{2\bar{d}^{1/3}}\bar{L}.   
\label{eq:bi}
\end{equation}

The speed ($v_{i}$) of the plasma flowing into the current sheet is derived by estimating the rate of change of flux below the current sheet, and is found to be:
\begin{equation}
\bar{v}_i = \frac{2}{3\bar{L}} \bar{d}^{-1/3},
\label{eq:vi}
\end{equation}
where $v_i$ is normalised with respect to  $v_0$.

\subsubsection{Energy Release}

The energy release follows by estimating the Poynting flux flow into the current sheet. 
The Poynting influx from both sides of the current sheet of length $L$ is:

\begin{equation}
    S_i = 2 \frac{v_{i}B_{i}^{2}}{\mu}L \pi z_s.
\label{eq:poynting}    
\end{equation}
The length of the current sheet depends on the type of reconnection. 
In the case for fast reconnection, $L$ is determined by assuming that the inflow speed has a known Mach \alfven number $\alpha$ ($v_{i}=\alpha v_{Ai}$, where $v_{Ai}=B_{i}/\sqrt{\mu \rho_i}$ and $\rho_i$ is the density of the inflowing plasma).
Then, using Equation (\ref{eq:bi}) and (\ref{eq:vi}), $L$ becomes
\begin{equation}
    \frac{L^{2}}{d_{0}^{2}}=\frac{4v_{0}}{9\alpha 
    v_{A0}}\frac{1}{[{1-(d/d_{0})^{4/3}}]^{1/2}},
\label{eq:lfast}
\end{equation}
where
\begin{equation}
    v_{A0}=\frac{B_{0}}{\sqrt{\mu \rho_i}},
\label{eq:hybrid}
\end{equation}
is a hybrid Alfv\'en speed. Then, from Equation (\ref{eq:poynting}), the Poynting flux into the current sheet for fast reconnection is
\begin{equation}
    \frac{\mu}{v_0 B_0^2 d_0^2} S_i= 2 \frac{2\pi}{3} \frac{M_{A0}}{\alpha}\frac{[1-\bar{d}^{4/3}]}{ \bar{d}^{2/3}}.
\label{eq:poyntingfast}
\end{equation}

Lastly, the energy release is derived by assuming that during fast reconnection, $\frac{2}{5}$ of the energy is converted to heat \citep{Priest_2014}:
\begin{equation}
    \frac{\mu}{v_0 B_0^2 d_0^2}  \frac{dW}{dt}=0.8 \frac{2\pi}{3} \frac{M_{A0}}{\alpha}\frac{[1-\bar{d}^{4/3}]}{ \bar{d}^{2/3}}.
\label{eq:heat}
\end{equation}

\section{Numerical Computations} 
\label{sec:numerical_simulations}

\subsection{Numerical Setup}
\label{sec:model}

We numerically solve the 3D MHD equations in Cartesian geometry using the Lare3D code (v3.2) \citep{Arber_etal2001}. The equations in dimensionless form are:
\begin{align}
&\frac{\partial \rho}{\partial t}+ \nabla \cdot (\rho \mathbf{v})  =0 ,\\
&\frac{\partial (\rho \mathbf{v})}{\partial t}  = - \nabla \cdot (\rho \mathbf{v v})  + (\nabla \times \mathbf{B}) \times \mathbf{B} - \nabla P + \rho \mathbf{g} , \\
&\frac{ \partial ( \rho \epsilon )}{\partial t} = - \nabla \cdot (\rho \epsilon \mathbf{v}) -P \nabla \cdot \mathbf{v}+ Q_\mathrm{j}+ Q_\mathrm{v} + Q_c, \\
&\frac{\partial \mathbf{B}}{\partial t} =\nabla \times (\mathbf{v}\times \mathbf{B})- \nabla \times( \eta \nabla \times \mathbf{B}),\\
&\epsilon  =\frac{P}{(\gamma -1)\rho}, \\
&P = \frac{\rho k_B T}{\mu_m},
\end{align}
where $\rho$, $\mathbf{v}$, $\mathbf{B}$ and $P$ are density, velocity vector, magnetic field vector and gas pressure. Gravity is $g_0=274$~m s$^{-1}$. Viscous heating ($Q_\mathrm{v}$) and Joule dissipation ($Q_\mathrm{j}$) are included. Heat conduction ($Q_c$) is treated using super-time stepping \citep{Meyer_etal2012}. We assume a perfect gas with specific heat of $\gamma=5/3$. The reduced mass is $\mu_m= m_f m_p$, where $m_p$ is the mass of proton and $m_f=1.2$. $k_B$ is the Boltzmann constant. 

In the 2D simulations of Paper II, we adopted a non-uniform resistivity profile that was a function of the current density. The functional form was selected to make the explicit resistivity  larger than  numerical diffusion. 
In the current 3D simulations, this is not possible, and so the resolution is such that  numerical diffusion dominates the reconnection region. Therefore, in this paper, we assume a low uniform explicit resistivity of $\eta=10^{-2}$ everywhere across the numerical domain and allow numerical diffusion to permit reconnection at the current sheet associated with flux cancellation. Exceptions to that are the boundaries, where $\eta$ decreases to zero so as to couple the field and flow properly there.

The normalisation is based on photospheric values of density $\rho_\mathrm{u}=1.67 \times 10^{-7}\ \mathrm{g}\ \mathrm{cm}^{-3}$, length-scale $H_\mathrm{u}=180 \ \mathrm{km}$ and magnetic field strength $B_\mathrm{u}=300 \ \mathrm{G}$. From these we obtain temperature $T_\mathrm{u}=6234~\mathrm{K}$, pressure $P_\mathrm{u}=7.16\times 10^3\ \mathrm{erg}\ \mathrm{cm}^{-3}$, velocity $v_\mathrm{u}=2.1\ \mathrm{km} \ \mathrm{s}^{-1}$ and time-scale $t_\mathrm{u}=86.9\ \mathrm{s}$.

\begin{figure}
\begin{center}
\includegraphics[width=\columnwidth]{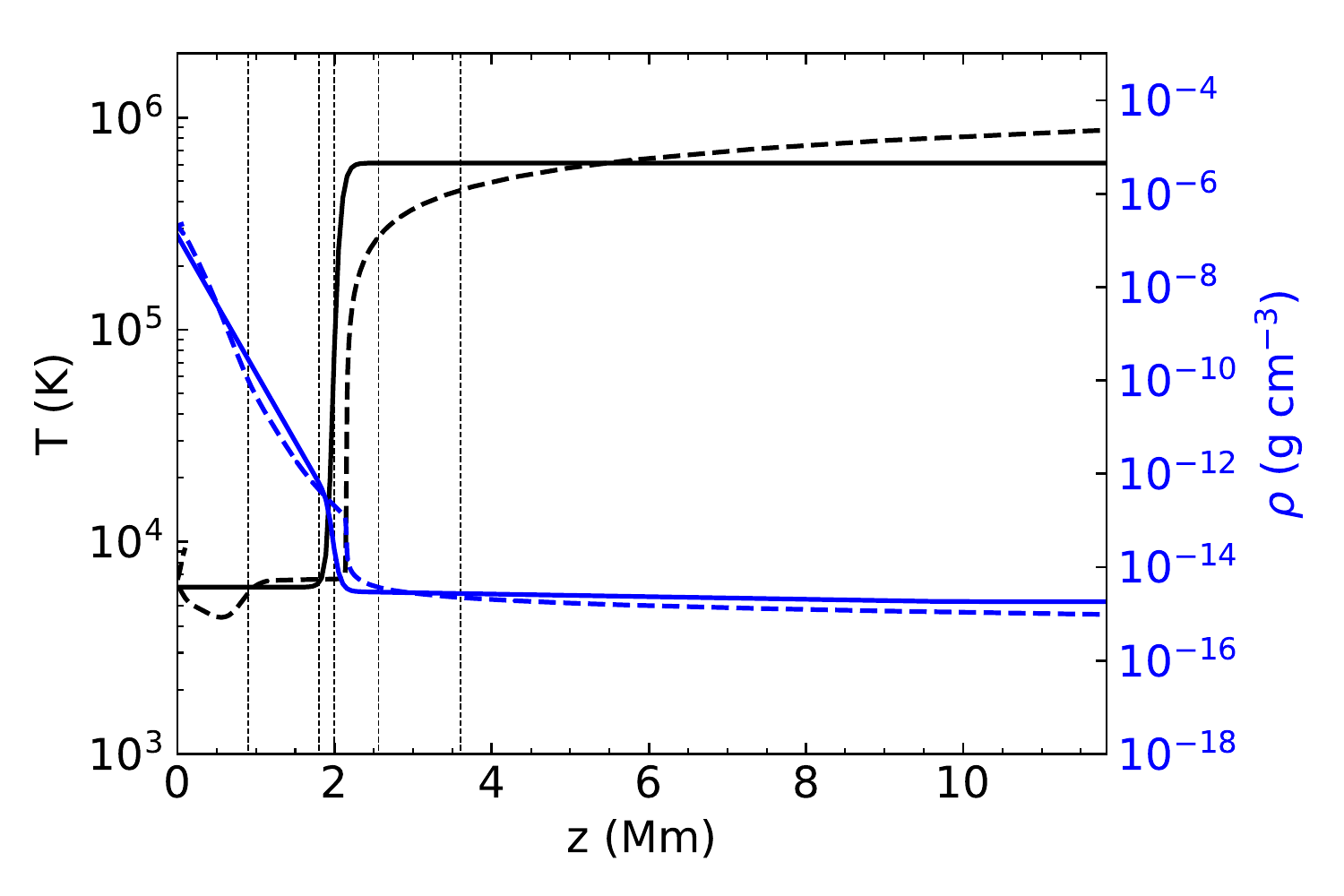}
\caption{
The atmospheric temperature (solid black) and density (solid blue). The dashed lines show the temperature and density of the 1D C7 model of \citet{Avrett_etal2008}. Vertical dashed lines show the heights of the null point at $t=0$ for  Cases 1-5 of Table~\ref{tab:table} (left to right, respectively).
}
\label{fig:stratification}
\end{center}
\end{figure}

The domain has a size of $x\in[-11.88,11.88]$~Mm and  $y\in[-11.88,11.88]$~Mm in the horizontal direction and $z\in[0,11.88]$~Mm in the vertical direction, on a $440\times440\times220$ uniform grid. A hyperbolic tangent profile is used for the atmospheric temperature, mimicking the steep temperature increase from the photosphere ($z=0$) to the corona:
\begin{align}
    T(z) = T_{ph} + \frac{T_{cor}- T_{ph}}{2} 
            \left( \tanh{\frac{z-z_{cor}}{w_{tr}} +1} \right),
\end{align}

where $T_{ph}=6109$~K, $T_{cor} = 0.61$~MK, $y_{cor}=2.12$~Mm and $w_{tr}=0.18$~Mm. 
This profile creates an isothermal photospheric-chromospheric layer at $0 \ \mathrm{Mm} \le z < 1.96 \ \mathrm{Mm} $, a transition region  at $1.9 \ \mathrm{Mm} \le z < 2.3 \ \mathrm{Mm}$ and an isothermal corona at $2.3 \ \mathrm{Mm} \le z < 11.88\ \mathrm{Mm}$.
The atmospheric density is derived by solving the hydrostatic equation $dP/dz = - gz$, assuming a photospheric density of $\rho_{ph}= 1.67\times10^{-7}$~g cm$^{-3}$.
The atmospheric temperature (solid black) and density  (solid blue) are shown in Figure~\ref{fig:stratification}. For comparison, we plot with dashed lines the temperature and density for the 1D model atmosphere (model C7) of \citet{Avrett_etal2008}.

\begin{figure}
\begin{center}
\includegraphics[width=\columnwidth]{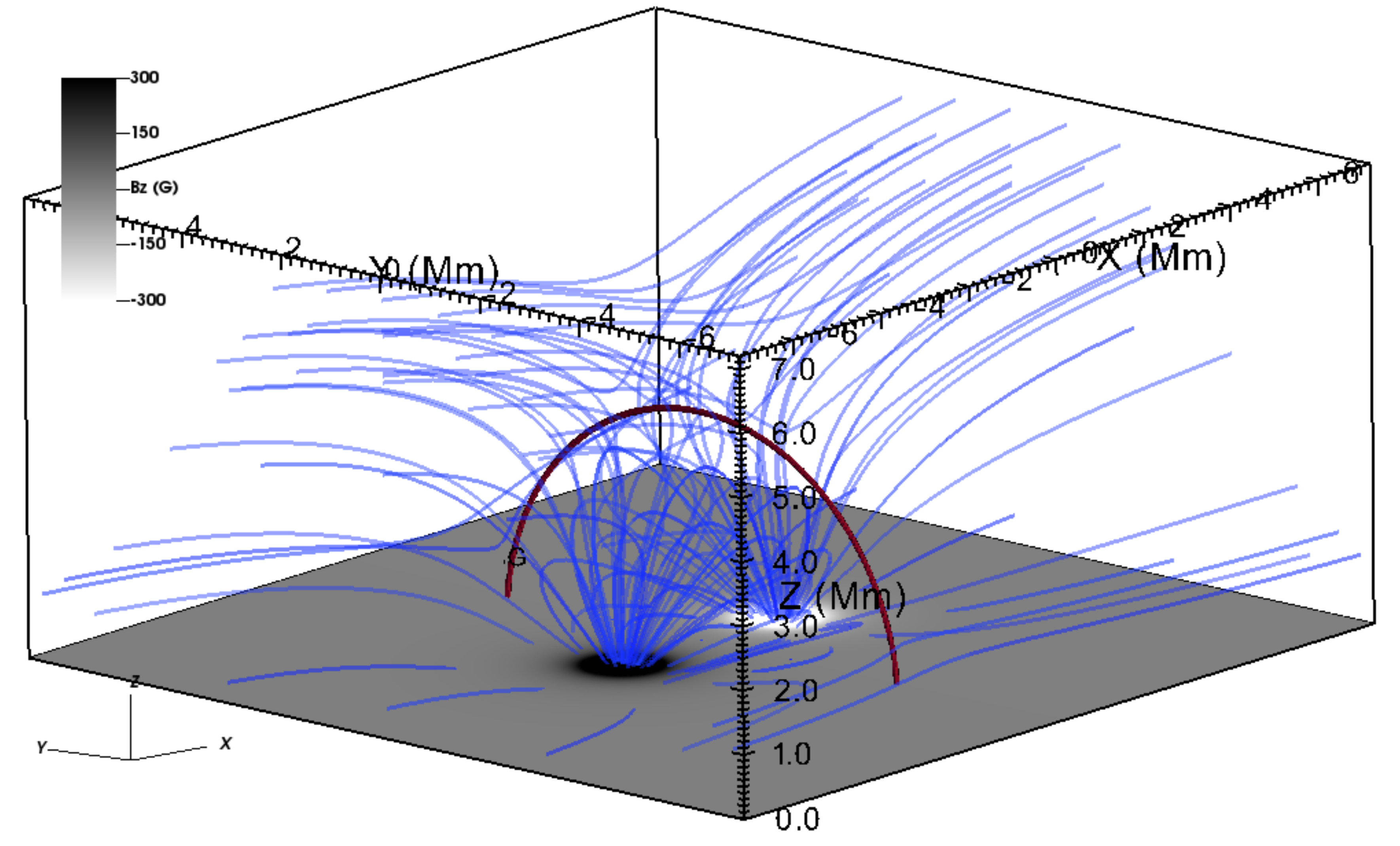}
\caption{
The 3D magnetic field configuration at $t=0$ (blue field lines) and the photospheric $B_z$ saturated at $\pm300~G$. Red contour of $|B|<0.5$~G, showing the location of a ring of null points.
}
\label{fig:magnetic_field3D}
\end{center}
\end{figure}

\begin{figure}
\begin{center}
\includegraphics[width=\columnwidth]{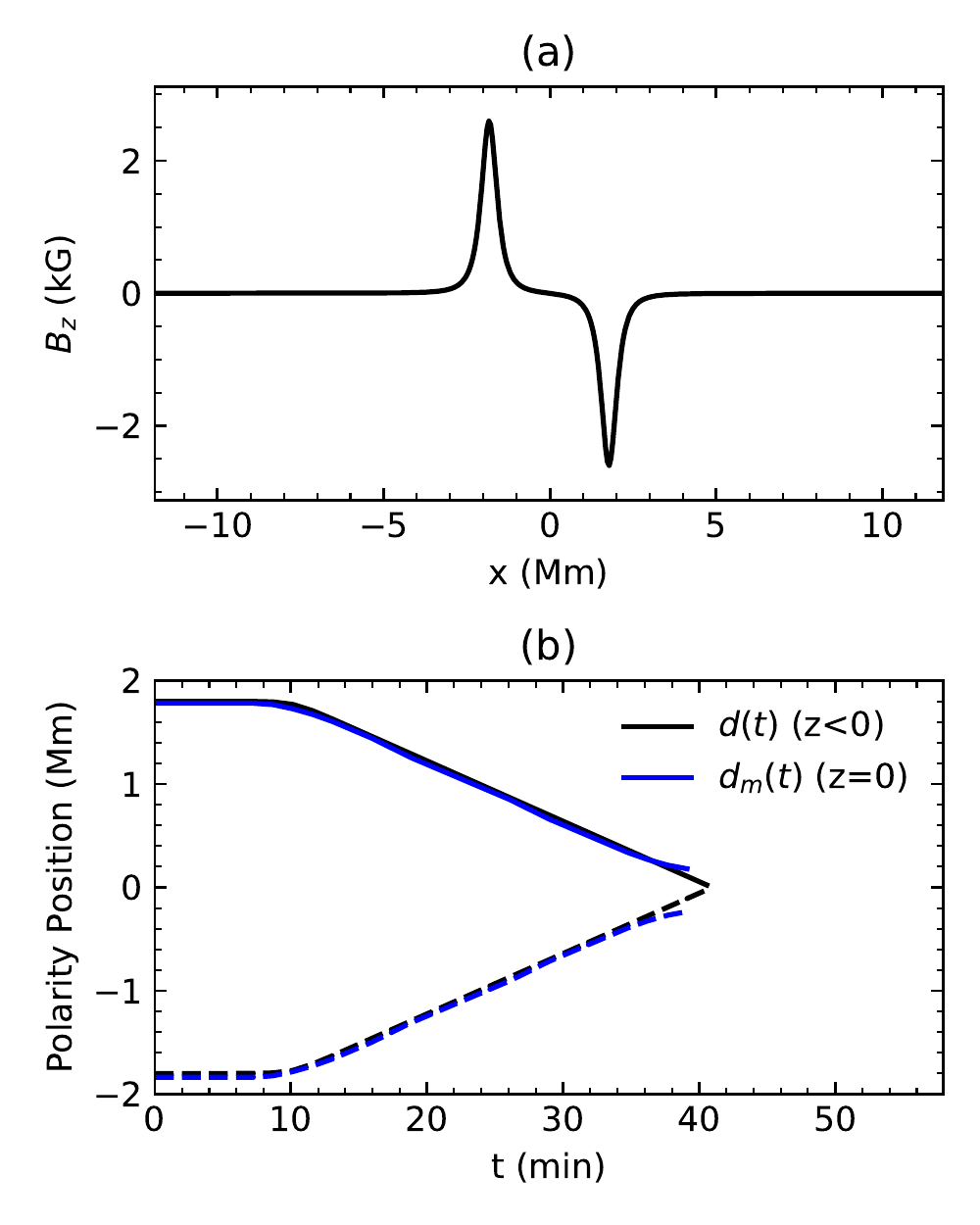}
\caption{
(a) Photospheric $B_z$ along the $x$-axis, at $y=0$. (b) The position of the sources ($d$, black lines) and the position of the photospheric polarities ($d_m$, blue lines) as a function of time.}
\label{fig:driver}
\end{center}
\end{figure}

The initial magnetic field is the sum of two magnetic sources and a horizontal field:
\begin{align}
    \mathbf{B} = \frac{F}{2\pi}  \frac{ \mathbf{\hat{r}_1} }{r_1^2 }  - \frac{F}{2\pi}  \frac{ \mathbf{\hat{r}_2} }{r_2^2} - B_0 \mathbf{\hat{x}},
    \label{eq:bfield_sim}
\end{align}
where
\begin{align}
    \mathbf{r_1} &= (x + d_s) \mathbf{\hat{x}} + y \mathbf{\hat{y}} + ( z - z_0) \mathbf{\hat{z}}, \\
    \mathbf{r_2} &= (x - d_s) \mathbf{\hat{x}} + y \mathbf{\hat{y}} + ( z - z_0) \mathbf{\hat{z}},
\end{align}
are the position vectors of the left and right sources, respectively, $d_s=1.8$~Mm is the distance of each source from the centre, $z_0=-0.36$~Mm is the depth of the sources below the photosphere (assumed outside  the numerical domain).
Each source has flux  $F=2.1\times10^{19}$~Mx, and the resulting photospheric polarities have a maximum field strength of 2.6~kG. The size of the polarities at the photosphere, defined as the length across which $|B_z|>100$~G, is 1.9 Mm (Figure~\ref{fig:driver}a).
The photospheric flux of each photospheric polarity is $F_{m} = 1.9\times10^{19}$~Mx.
The horizontal field has a strength of $B_0= 15$~G (case 1).
The magnetic field configuration at $t=0$ is visualised in Figure~\ref{fig:magnetic_field3D}. Other cases of $B_0$ were also examined, where $B_0$ was varied in order to vary the initial height of the null points in the atmosphere. 
The values of $B_0$ and the corresponding maximum height of the null point at $x=y=0$ at $t=0$~min are shown in Table~\ref{tab:table}.

The boundary conditions are imposed in a similar way to  Paper II. At the lower boundary (photosphere), the density and energy are assumed to have zero gradient.
The simulation is driven by changing the magnetic field at the lower boundary using Equation~(\ref{eq:bfield_sim}) and varying the source positions $\pm d(t)$ from their initial values $\pm d_s$ according to $d(t) = d_s - x(t)$, where $x(t)$ is
\begin{eqnarray}
    x(t) = v_{max} \frac{w}{2} \left[  
                                      \ln{ \left( \cosh{ \frac{t-t_0}{w} } \right)} -
                                      \ln{ \left( \cosh{ \frac{t_0}{w} } \right)} 
                                \right]\\ \nonumber +
          \half v_{max} t, \ \ \ \ \ \ \ \ \ \ \ \ \ \ \ \ \ \ \ \ \ \ \ \ \ \ \ 
\end{eqnarray}
and $v_{max}$ = 1 \kms, $t_0=10.1$~min and $w=1.4$~min. This $x(t)$ leads to source velocities of $\pm v_0(t)$, where
\begin{align}
    v_0(t) = {\half v_{max}}  \left( \tanh{ \frac{t-t_0}{w} } +1  \right).
\end{align}
In Figure~\ref{fig:driver}b (black line) we show the half-separation ($d(t)$) of the sources (below the photosphere, outside the numerical domain) as a function of time. The blue lines show the positions of the polarities at $z=0$ (found by measuring the location of maximum $B_z$ at the photosphere). The latter reflects the response of the photosphere to the driver.
At the rest of the boundaries, we assume $\mathbf{v}=0$ and zero gradients for $\mathbf{B}$, $\rho$, $\epsilon$.

\begin{deluxetable}{ccCrlc}[t]
\tablecaption{Initial Conditions for the Simulations \label{tab:table}}
\tablecolumns{3}
\tablenum{1}
\tablewidth{0pt}
\tablehead{
\colhead{Name} &
\colhead{$B_0$ (G)} &
\colhead{$z_s$ (Mm)}
}
\startdata
Case 1 &    15     & 3.6  & \\
Case 2 &    30     & 2.6  & \\
Case 3 &    47     & 2  & \\
Case 4 &    53     & 1.8  & \\
Case 5 &    114    & 0.9  & \\
\enddata
\end{deluxetable}

\subsection{Comparison of Theory with Simulation}
\label{sec:theory_comparison}

In this section, we discuss case 1 of Table~\ref{tab:table} and use it to compare the simulation with the analytical theory summarised in Sec.~\ref{sec:theory}.

\subsubsection{Brief Description of Simulation}

\begin{figure}
\begin{center}
\includegraphics[width=\columnwidth]{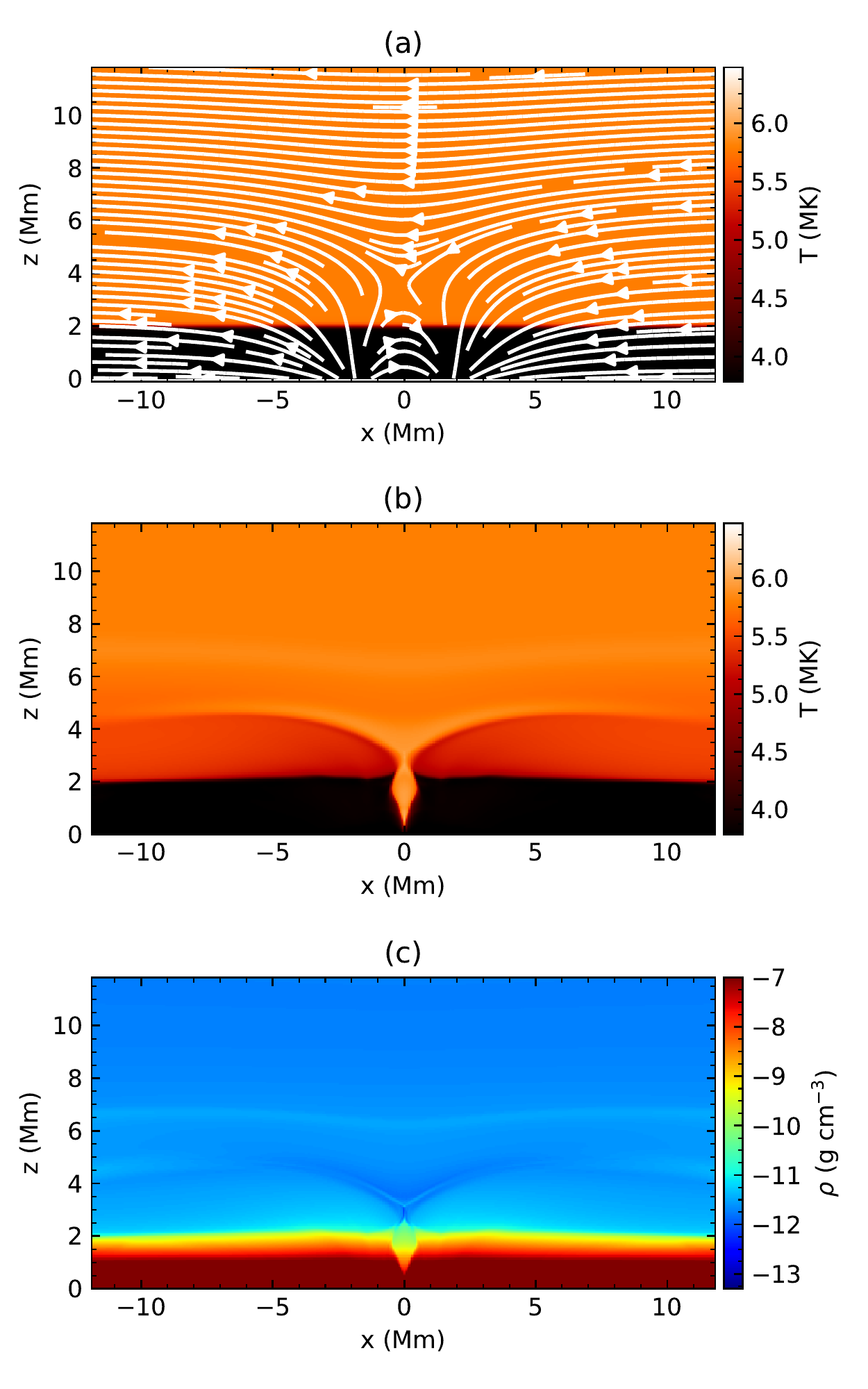}
\caption{ 
(a) Temperature and magnetic field at $t=0$ at the $xz$-midplane.
(b) Temperature and (c) density at $t=43$~min at the $xz$-midplane.
}
\label{fig:temp_rho}
\end{center}
\end{figure}

\begin{figure}
\begin{center}
\includegraphics[width=\columnwidth]{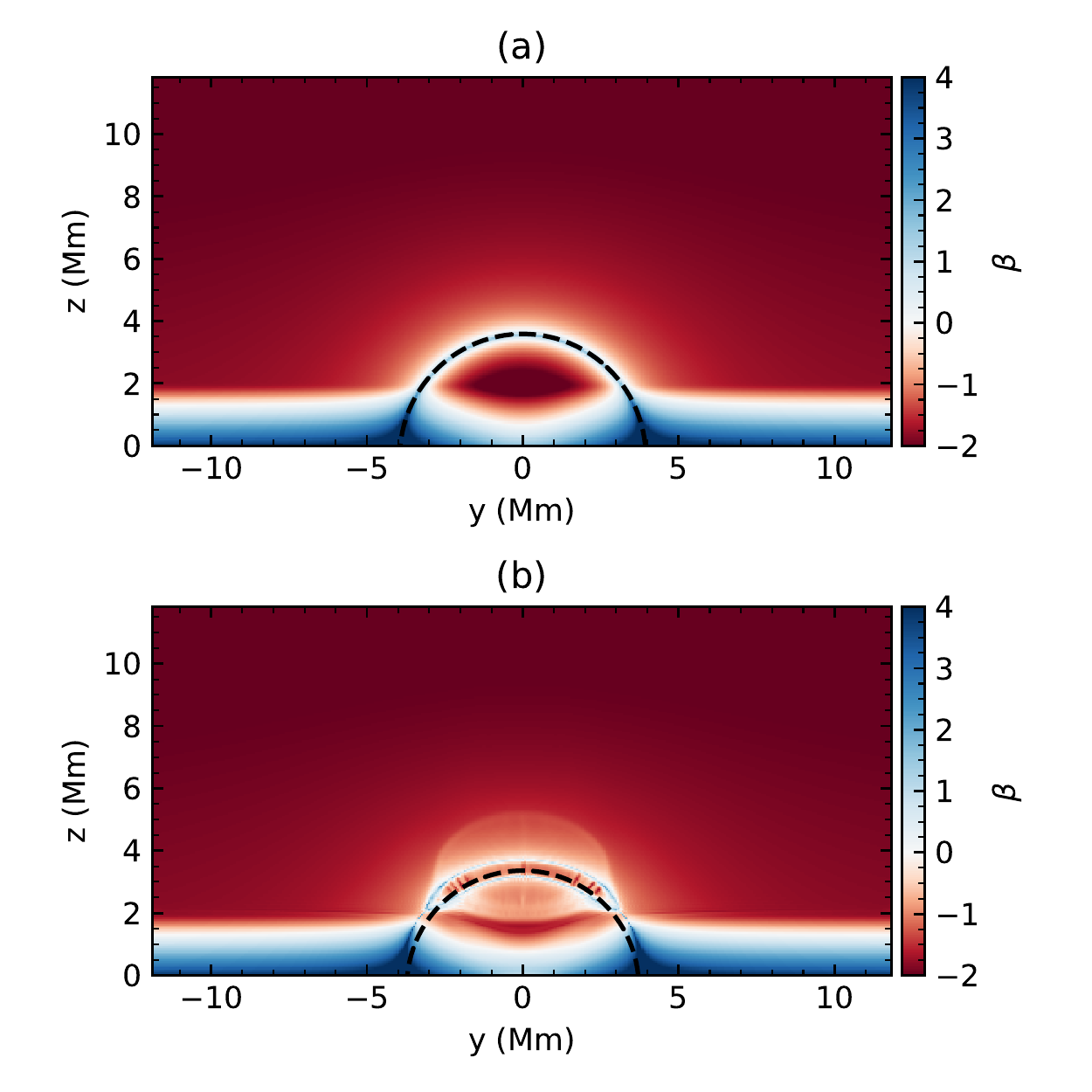}
\caption{ 
Plasma $\beta$ at the $yz$-midplane ($x=0$) at (a) $t=0$ and (b) $t=17.4$~min. The dashed lines show the location of the null points according to Equation~\ref{eqn11}.
}
\label{fig:beta}
\end{center}
\end{figure}

The 3D magnetic field at $t=0$ (Figure \ref{fig:magnetic_field3D}) shows that, above the magnetic arcade, a semicircular ring of null points (red contour) extends from the photosphere and reaches a maximum height at the $xz$-midplane ($y=0$). 
At that plane, the  null is located in the corona at a height of $(x,z)=(0,3.6)$~Mm (Figure~\ref{fig:temp_rho}a). To further visualise the ring of null points, in Figure~\ref{fig:beta} we show the plasma $\beta$ at the $yz$-midplane, and the positions of the null point according to Equation~\ref{eqn11} (dashed line). 
Because the background atmosphere is stratified, the null points are located both inside regions of low plasma $\beta$ (between $y=\pm3.5$~Mm, red colour around dashed line) and of high plasma $\beta$ (blue colour around dashed line). 

When the driver is switched on, reconnection is not driven across the whole semicircular ring of null points because of the different plasma $\beta$ environments. 
Instead, the converging photospheric polarities drive reconnection only between $y=\pm3.5$~Mm (Figure~\ref{fig:beta}b), which needs to be taken into account when comparing the theory with the simulation, and will be discussed later.

The energy released by reconnection heats locally the plasma, which spreads above and below the null points (and shows up as a ``horizontal'' heated region and an underlying heated arcade in Figure~\ref{fig:temp_rho}b), and is denser than the background atmosphere (Figure~\ref{fig:temp_rho}c).

After $t=36$~min, the photospheric polarities ($d_m(t)$, blue lines, Figure~\ref{fig:driver}b) do not keep following the driver (black lines). This is  because the magnitude of the photospheric field has decreased to the point that $\beta>1$. As a result, the driver cannot move the overlying field anymore, and therefore, the reconnection at the null points gradually stops. Above the photosphere, there is still a remaining magnetic arcade.

\subsubsection{Comparison Methodology}

When deriving the theory, we assumed that the converging polarities drive reconnection across the whole semicircular ring of null points (i.e., across the dashed line of Figure \ref{fig:beta}a). Therefore, we assumed that the length along which reconnection occurs is  $\pi z_S$. 
In the simulation, however, reconnection is not driven across the whole semicircular ring because of the different plasma $\beta$ regions.
Instead, the converging photospheric polarities drive reconnection only along a dashed line segment between $y=\pm3.5$~Mm (Figure~\ref{fig:beta}b). We call the length of that line segment $l$. 
To compare theory and simulation, the theory has to  be adjusted to predict the energy released associated with the length $l$ rather than $\pi z_S$. To do so, we multiply Equations (\ref{eq:poyntingfast}) and (\ref{eq:heat}) by a correction factor of $l / (\pi z_S)$.

In Paper II, we compared the simulation with the theory using two approaches. 
The first was to make the theoretical estimates using the parameters characterising the driver (e.g.,  the flux and time-distance profile of the sources below the photosphere).
The second approach was to measure the photospheric and atmospheric response to the driver (e.g., measured photospheric flux and time-distance profile of the photospheric polarities) and use these with the theoretical expressions. It was found that the latter approach, which mimics an actual observation, gave a better comparison between simulation and theory, and so that is the approach we adopt here.

\begin{figure}
\begin{center}
\includegraphics[width=\columnwidth]{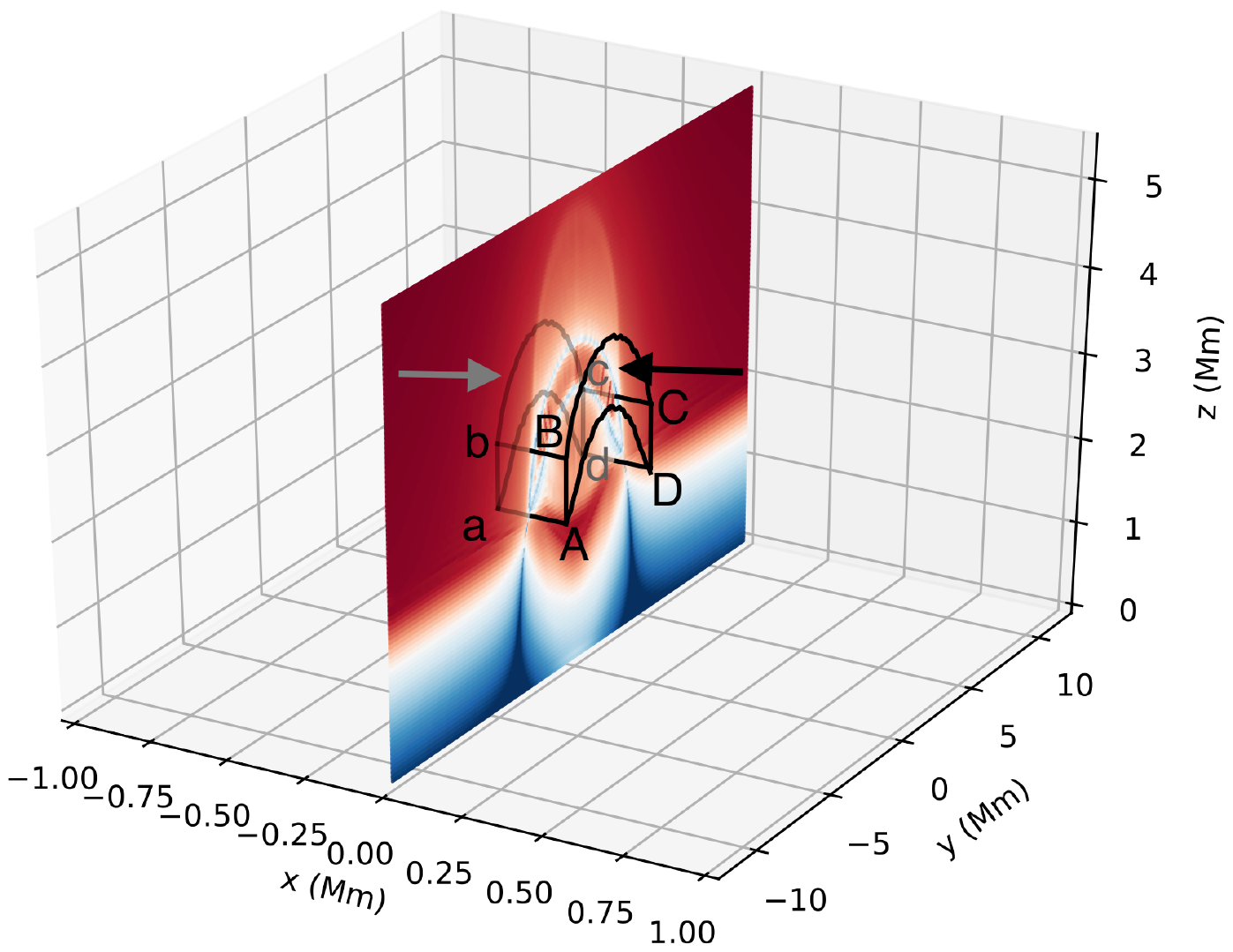}
\caption{
    Three-dimensional representation of the characteristic surface inside which the current sheet is located at $t=17.4$~min (solid  lines). The red-blue plane is the same as in Figure~\ref{fig:beta}b. Arrows show  the sides through which plasma inflows into the current sheet.
}
\label{fig:cs_3d}
\end{center}
\end{figure}

To compare  theory with  simulation, we need to identify the characteristic surface around the current sheet as it evolves in time.
The current sheet is a structure located approximately between $y=\pm3.5$~Mm along the $y$-axis. 
At each point along the $y$-axis, we identify the vertical extent along the $z$-axis, tracing the arced shape of the current sheet (the distance between the between blue lines around $y=\pm3.5$~Mm, Figure \ref{fig:beta}b). 
To visualise this shape, in Figure~\ref{fig:cs_3d} we plot the $yz$-midplane of Figure \ref{fig:beta}b in a three-dimensional volume. The traced arced black lines (e.g., AD and BC) follow the shape of the current sheet. 
We then identify the regions that are parallel to the current sheet and at a distance of $\Delta x=0.2$~Mm away from it (ABCD and abcd surfaces). 
This distance is such that the current density there is at least an order of magnitude lower than the one inside the current sheet.
Using these two surfaces, we identify a three-dimensional surface, delineated  by the black lines, inside which the current sheet is located. This shape is changing over time as the polarities converge.

We compare the theory with the simulation as follows. 
The length ($L$) of the current sheet  is taken for simplicity to be the vertical extent of the current sheet along the $z$-axis at $x=y=0$. We do so since the length of the current sheet is approximately radially symmetric (see Figure \ref{fig:beta}b).
The values of the inflowing magnetic field strength ($B_i$), inflowing velocity magnitude ($v_i$) and inflowing density ($\rho_i$) are taken as the average values at the surfaces parallel to the current sheet (ABCD and abcd, Figure \ref{fig:cs_3d}). 
The total inflow of Poynting flux ($S_{i}$) into the current sheet is measured by taking into account the sum of the Poynting flux through both the abcd and ABCD surfaces.

\begin{figure*}
\begin{center}
\includegraphics[width=\textwidth]{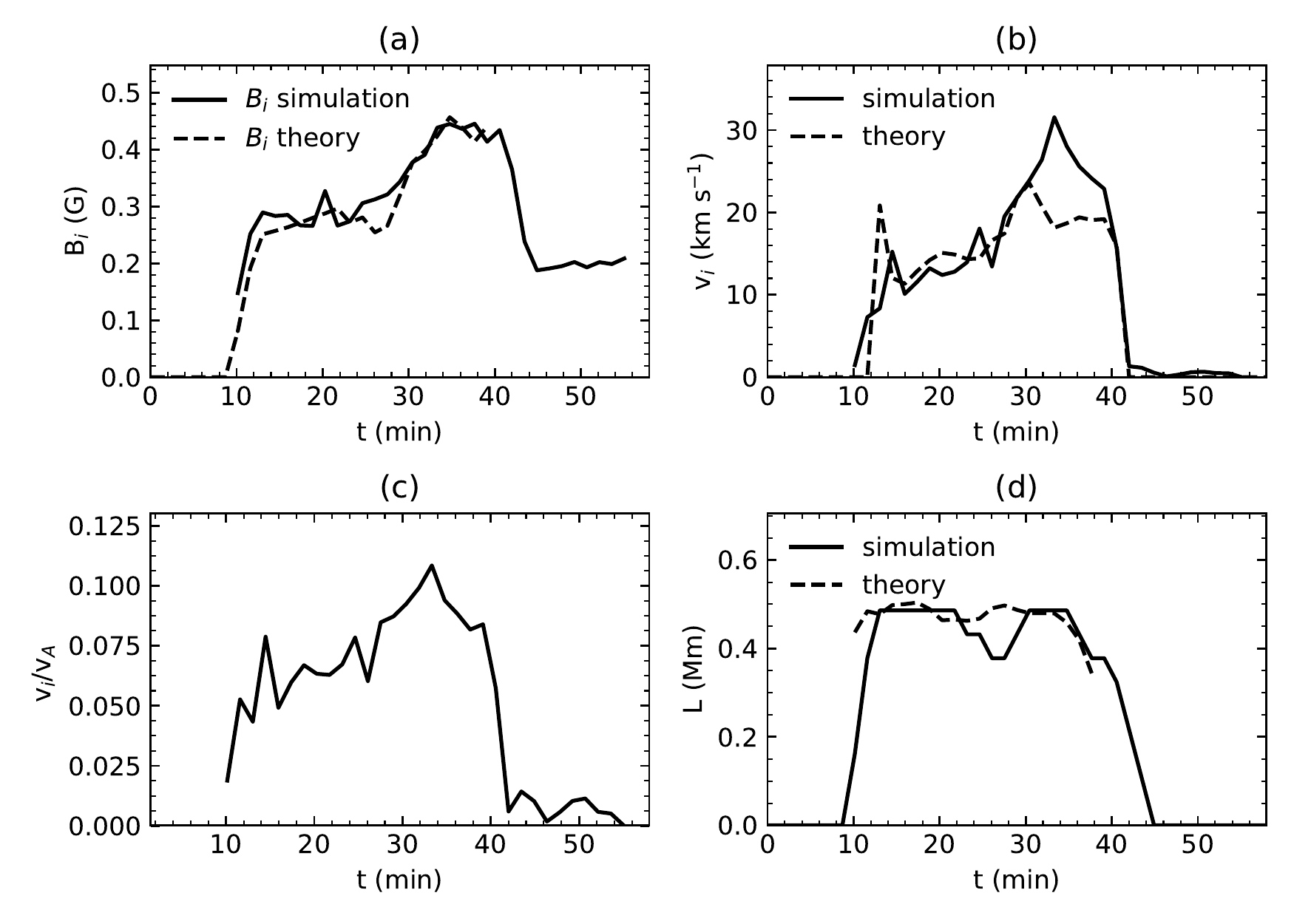}
\caption{
Comparison between simulation and theory for 
a) the inflow magnetic field,
(b) the inflow velocity,
(c) the  \alfven Mach number of the inflow,
(d) the length of the current sheet.}
\label{fig:inflow_var}
\end{center}
\end{figure*}

\subsubsection{Comparison of Theory and Simulation}

The inflow magnetic field strength ($B_i$) and velocity magnitude ($v_i$) are plotted in Figure~\ref{fig:inflow_var}a,b. The solid lines are measured from the simulation and the dashed lines are the theoretical estimates using Equations \ref{eq:bi} and \ref{eq:vi} respectively. We see that the theory is in good agreement with the simulation.

To estimate the length of the current sheet using Equation \ref{eq:lfast} we have to measure $\alpha$ (\alfven Mach number) and  $v_{A0}$ using Equation~(\ref{eq:hybrid}). For $\alpha$, we use average value of the \alfven Mach number between $t=10$ min and $40$~min, during which the cancellation occurs, which is $\alpha=0.075$ (Figure~\ref{fig:inflow_var}c). This value of $\alpha$ is typical for fast reconnection \citep{Priest_2014}.
Figure~\ref{fig:inflow_var}d shows the length of the current sheet from the simulation (solid line) and from the theoretical estimate (dashed line). Both are in good agreement.

\begin{figure}
\begin{center}
\includegraphics[width=\columnwidth]{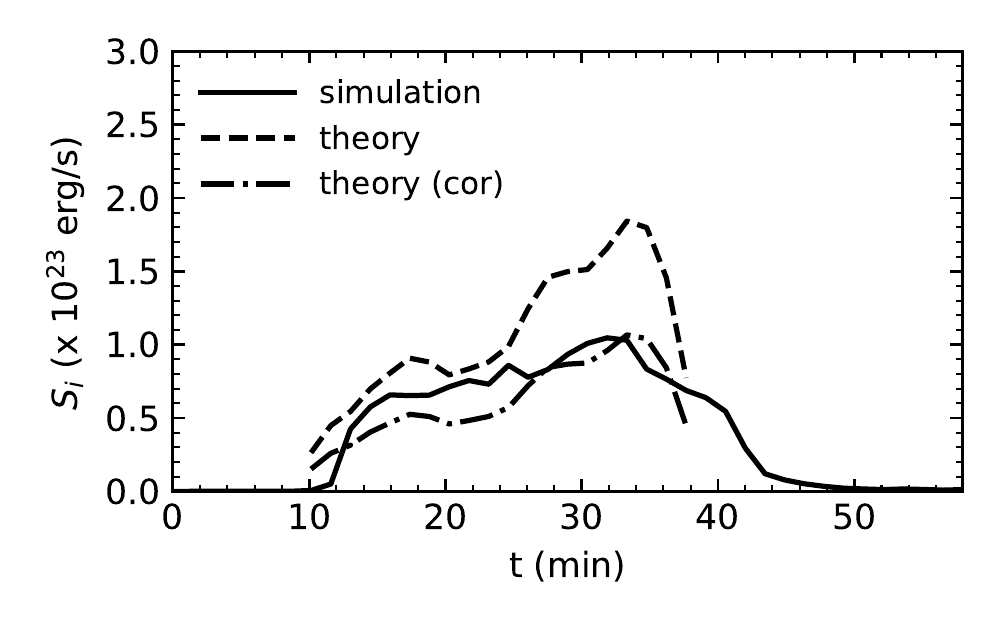}
\caption{ 
Comparison between the simulation, the theory and the ``corrected'' theory for the total  Poynting influx.
}
\label{fig:energy}
\end{center}
\end{figure}

We now estimate the Poynting flux  into the current sheet. The measured Poynting inflow is shown in Figure \ref{fig:energy} (solid line). The dashed line shows the analytical estimate using Equation (\ref{eq:poyntingfast}). The  analytical estimate becomes larger than the measured Poynting flux by up to a factor of 2. 
This is because Equation (\ref{eq:poyntingfast}) assumes that the length along which reconnection occurs is $\pi z_S$. However, as discussed previously, in the simulation, reconnection occurs only along a part of the semicircular ring of null points with length $l$. If we correct the theoretical prediction by taking this into account  and multiply Equation (\ref{eq:poynting}) by a correction factor of $l/(\pi z_S)$, then the theory is in good agreement with the simulation (dot-dashed line). 
Similar agreement is found for the other cases of Table~\ref{tab:table}.

In Paper II, we also compared the conversion of Poynting flux to kinetic and thermal energy. To do a similar analysis for the 3D simulation, we would have to calculate the energy integral term $-\int_V \eta \mathbf{j}^2 dV$ and compare it with the Equation~(\ref{eq:heat}). 
However, as discussed in Sec.~\ref{sec:numerical_simulations}, it is numerical resistivity that drives the reconnection at the current sheet. Therefore the explicit value of $\eta$ is unknown, and so we do not here compare the analytical expression for the release of heat with the simulation. We note however, that the Poynting inflow is estimated well by the theory, as also are the parameters associated with the reconnection. So, we expect the expression for the conversion of Poynting flux inflow to heat to be also well estimated, since it is simply a proportion of the Poynting  inflow.

\begin{figure*}
\begin{center}
\includegraphics[width=\textwidth]{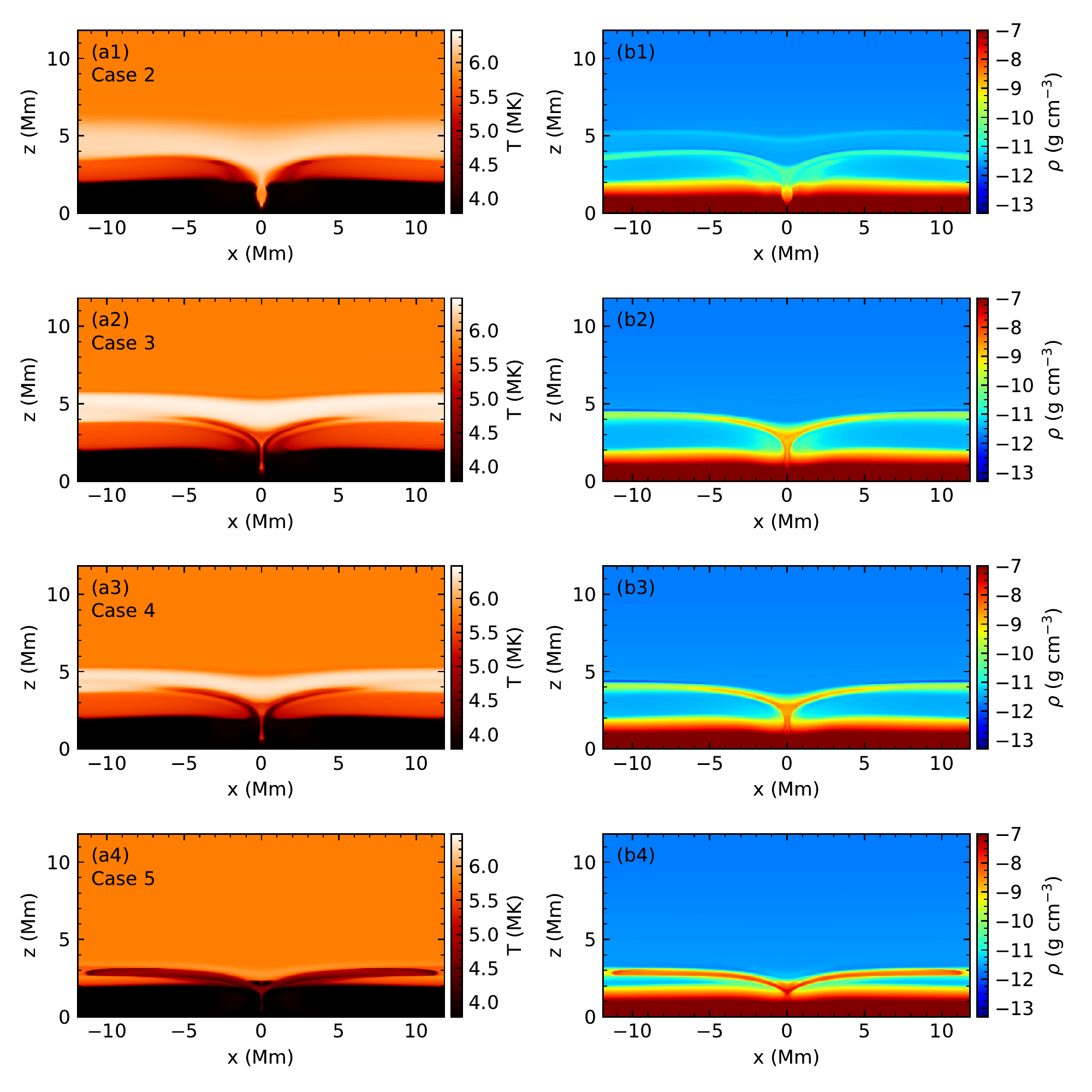}
\caption{ Temperature (left column) and density (right column) for cases 2-5 at  $y=0$ for $t=43$~min.}
\label{fig:parametric}
\end{center}
\end{figure*}

\subsection{Atmospheric Response}
\label{sec:parametric}

We now discuss the atmospheric response following  reconnection. For this we vary the atmospheric field strength $B_0$ in order to change the initial maximum height of the nulls at $t=0$, thus initiating reconnection at different atmospheric heights. 
The initial maximum height of the nulls for different values of $B_0$ is shown in Table~\ref{tab:table}. 
These cases are similar to the ones used in Paper II, and are selected so that the initial maximum height of the nulls is placed either high in the corona (case 1), at base of the corona (case 2), at the middle of the transition region (case 3), at the base of transition region region (case 4) or near the photosphere (case 5). These heights are shown against the initial stratification in Figure~\ref{fig:stratification} (as vertical lines).

In Paper II we found that, during the convergence of the polarities, starting at its maximum height, the current sheet moved towards the photosphere, and thus the reconnection occurred at progressively lower atmospheric heights. 
When the reconnection occurred in the corona or upper transition region, the resulting outflow was hot. In contrast, when the reconnection occurred at lower heights, a cool outflow was also produced. Therefore, depending on the  height of the null point, either hot ejections or cool ones or a combination of both hot and cool ejections were formed. These hot and cool outflows were formed with or without a time difference, depending on the initial conditions.
The same qualitative behaviour is found in the 3D simulations, with the addition of spatial effects due to the 3D extent of the current sheet.

\begin{figure*}
\begin{center}
\includegraphics[width=\textwidth]{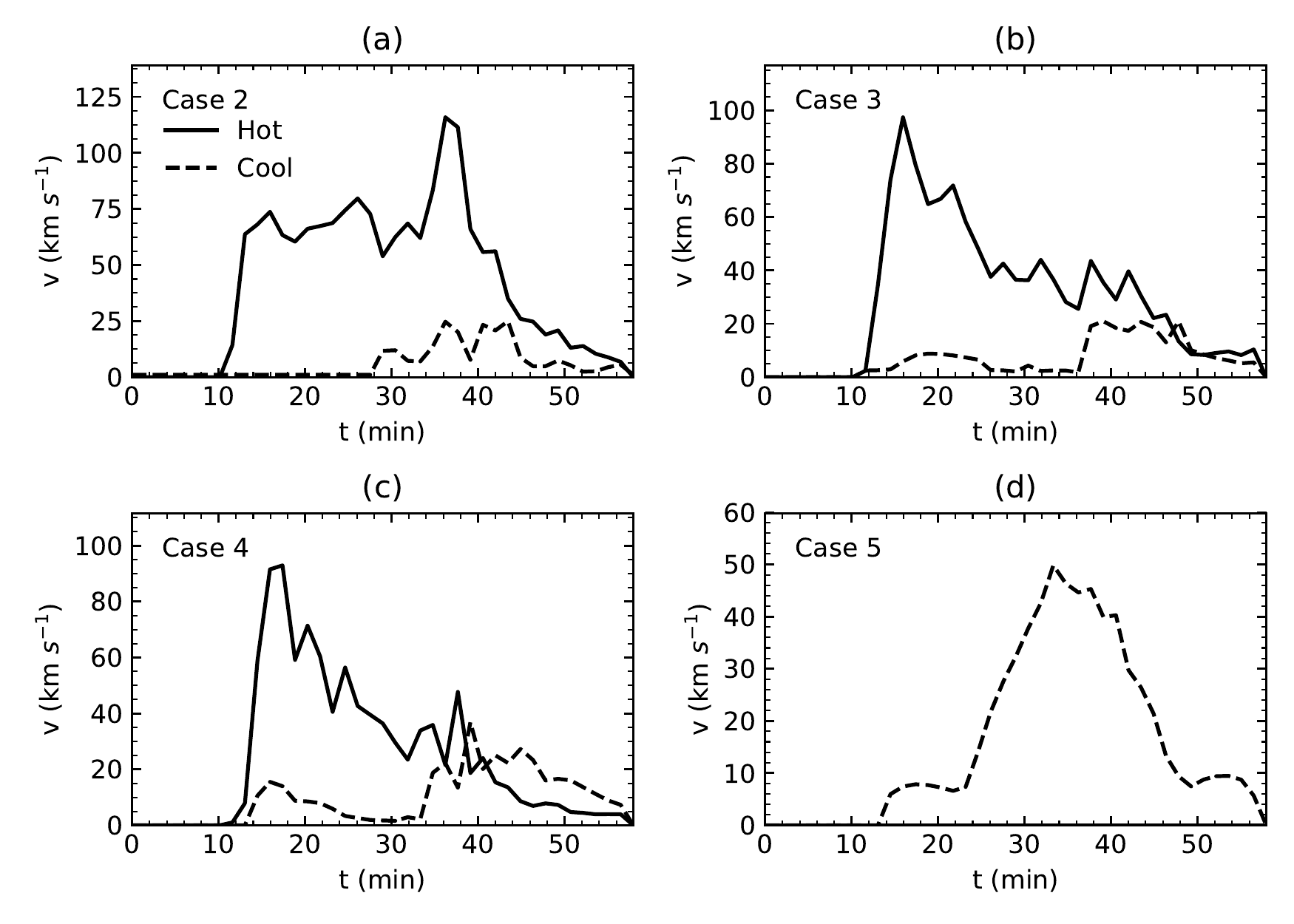}
\caption{ Maximum velocity of the hot ($T>1$~MK, solid lines) and cool ($T<0.2$~MK, dashed lines) plasma, measured at the $y=0$ Mm plane, for (a) Case 2, (b) Case 3, (c) Case 4 and (d) Case 5.}
\label{fig:fig_outflow_speeds}
\end{center}
\end{figure*}

Figure~\ref{fig:parametric} shows the temperature and density for cases 2 to 5 at the $y=0$ midplane after the driving has stopped. 
(Case 1 is shown in Figure~\ref{fig:temp_rho}.) Depending on the height of the null, the resulting outflows can be either hot or cool  or can consist of a combination of both hot and cool plasmas. 
By examining the time evolution of the maximum velocity of the hot  ($T>1$~MK) and cool ($T<0.2$~MK) outflows at that plane (Figure~\ref{fig:fig_outflow_speeds}), we find that, when the null point is initially located higher in the atmosphere, the hot ejection is produced first ($t=12$~min), and the cool ejection is produced later ($t=30$~min) (panels (a)). 
For a null initially located lower in the atmosphere, the cool and the hot ejections have less temporal separation (cool ejections appear from $t=16$~min and $t=12$~min, (b) and (c)). For a null close to the photosphere, only a cool ejection is formed (d).
These results are qualitatively similar to those of the 2D simulations in Paper II.

However, in 3D, instead of having only one null point at one height, we have a ring of nulls (or, more generally, a separator), and so the convergence of the polarities drives reconnection simultaneously at many different heights (e.g., Figure \ref{fig:beta}b).
Therefore, reconnection can  occur at different heights not only because the current sheet moves downwards during the convergence of the polarities, but also because of the spread of nulls or a separator over a range of  heights.
This produces both hot and cool ejections with a spatial offset.

\begin{figure*}
\begin{center}
\includegraphics[width=\textwidth]{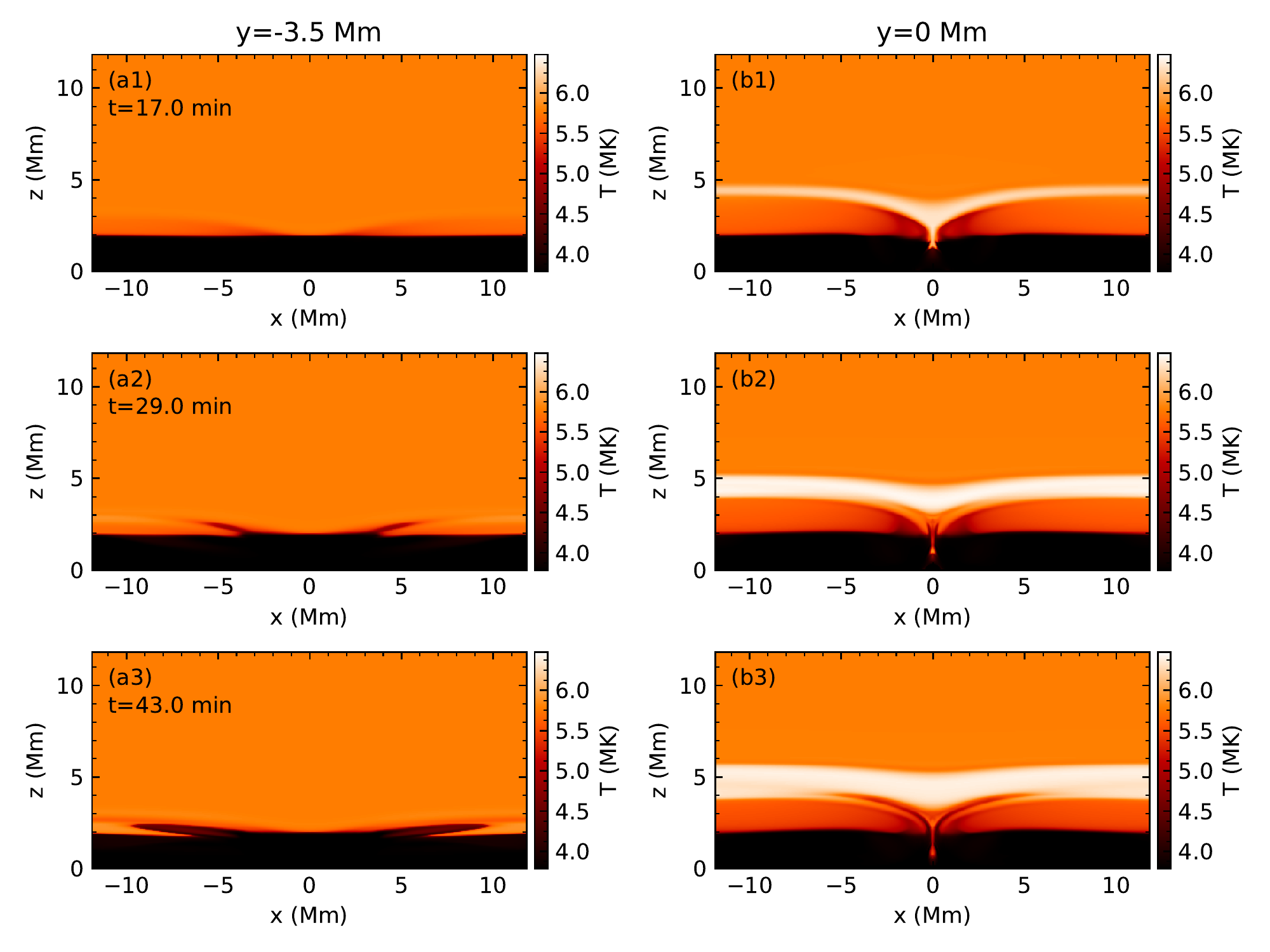}
\caption{
Temperature at the $y=-3.5$~Mm (left column) and the $y=0$~Mm (right column) plane for case 3 at  $t=17$~min (first row), $t=29$~min (second row) and $t=43$~min (third row).
}
\label{fig:parametric_offset}
\end{center}
\end{figure*}

To visualise the above, in Figure~\ref{fig:parametric_offset} we show for case 3 the temperature at an offset plane ($y=-3.5$~Mm) and at the midplane of the numerical domain ($y=0$~Mm), at three different times ($t=19$~min, first row; $t=29$~min, second row; $t=43$~min, third row).
At the $y=-3.5$~Mm plane, the null at $t=0$ is located lower in the atmosphere in comparison to the null at the $y=0$~Mm plane. 
Therefore, at that plane the convergence of the polarities results in  mostly cool ejections (panels (a1), (a2), (a3)).
At the same time, at the $y=0$ plane, the ejections are initially hot (panels (b1), (b2)), while cool ejections form with a time delay as the null progressively moves lower (panel (b3)). 
The latter cool outflows are spatially offset in comparison to the $y=-3.5$~Mm ones and develop at different times. In addition, panels (a1), (b1) show that, during the initial stages of the energy release, the hot ejection at the $y=0$~Mm plane is formed sooner than the offset cool one at the $y=-3.5$~Mm plane. This is because more energy is needed to accelerate upwards the denser material located lower in the atmosphere.

\begin{figure*}
\begin{center}
\includegraphics[width=\textwidth]{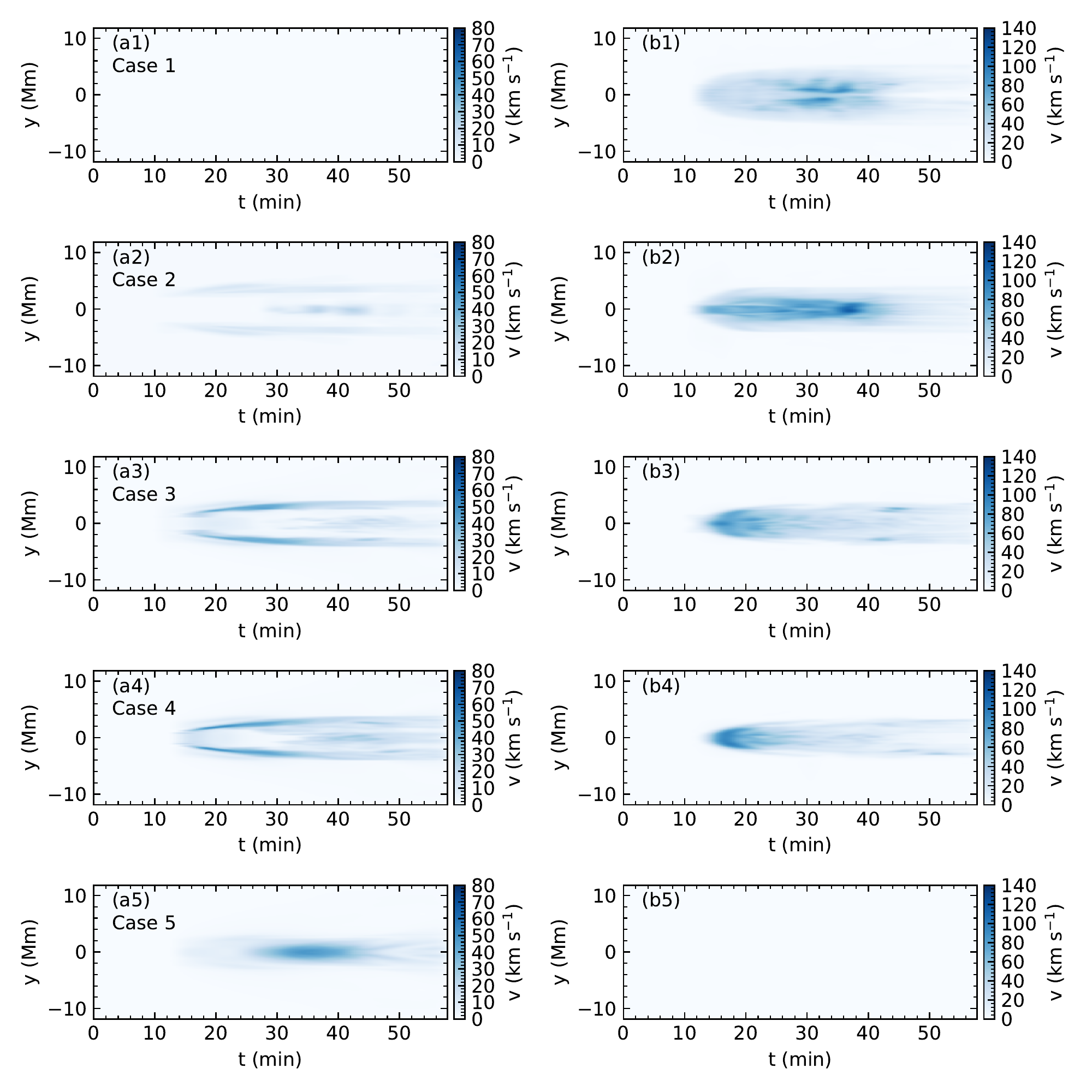}
\caption{ 
    2D maps of the time evolution of the maximum velocity at each $y$-plane for the cool ($T<0.2$~MK, left column) and the hot ($T>1$~MK, right column) plasma components, for all the different cases (rows).
}
\label{fig:fig_outflow_speeds_offset}
\end{center}
\end{figure*}

We further demonstrate these spatial effects by computing 2D maps of the temporal evolution of the maximum velocity of the hot  ($T>1$~MK) and cool ($T<0.2$~MK) outflows at all $y$-planes (Figure~\ref{fig:fig_outflow_speeds_offset}).
In Case 1 we do not find any cool component as the reconnection occurs mainly in the corona. In Case 2, we find the first signs of  spatially offset hot and cool flows. At $t$=14~min, the hot outflow originates from around $y=0$, whereas the two offset cool outflows originate from around $y\approx\pm3.5$~Mm. Throughout the cancellation, these cool outflows are at different spatial locations from the hot outflow. At $t>30$~min another cool outflow starts to appear around $y=0$  since the null point at that plane has now moved lower. This flow is co-spatial with the hot outflow originating from a similar location.
A gap appears between the cool outflows in panel (a2) around $t=30-45$ min.  Asymmetric shrinking of the ring of null due to  differences in $\beta$ causes a pressure gradient below the ring of nulls, which lifts the dense atmosphere around $y=0$ upwards. Therefore, at $y=0$, the dense atmosphere is higher than at $y\approx2$, and so reconnection at $y=0$ produces a cooler outflow than at $y\approx2$, so creating the gap.
In Cases 3 and 4, the cool outflows are again spatially offset with their magnitudes becoming stronger as the nulls are located lower in the atmosphere. In addition, the cool outflows around $y=0$ start to appear earlier, as in Paper II. In Case 5, we  find a cool outflow originating only from the central part of the cancelling region.

Adding to our findings from Paper II, our results indicate that, depending on the height of the null point and the parameters of the cancellation, during flux cancellation: i) hot ejections or cool ejections or a combination of both hot and cool ejections can be formed, ii) these can be formed with or without a time difference and iii) these can be formed with or without a spatial offset.

\section{Discussion}
\label{sec:discussion}

In Paper I we presented a theoretical model on how magnetic reconnection driven by photospheric flux cancellation can act as a mechanism for energizing coronal loops and heating the chromosphere. In Paper II we numerically validated our theoretical estimates by the means of 2D simulations of two converging polarities inside a stratified atmosphere with a horizontal background field. In the present work, we further the numerical validation and study the atmospheric response to the cancellation by performing 3D simulation.

In our theoretical model we assumed that as the two polarities converge, reconnection is driven at a circular ring of null points that extends from the photosphere up to a maximum height in the atmosphere. 
In the simulations, because the field is inside a stratified atmosphere, the convergence of the field does not drive reconnection along the whole length of that circular ring. It does so, only along the length of the ring of nulls that is located in a low $\beta$ environment. The part of the ring of nulls located in a high $\beta$ environment does not react to the convergence of the polarities.
When comparing the theoretical model to the 3D simulation, we found that the simulated total Poynting flux flowing into the current sheet differed from the theoretical estimate by a factor of $\sim2$. After correcting the theoretical estimate to account for the fact that only a part of the ring of nulls is involved in the reconnection, the theory was in excellent agreement with the simulation.

In Paper II, besides the Poynting influx, we compared the energy release rate from the simulation with the theory. We do not make such a comparison in the current paper, since the numerical resistivity dominates. 
However, all the quantities involved in the analytical estimation of the rate of energy release (inflowing magnetic field strength and velocity, current sheet length, Poynting influx) are in agreement with the simulation. 
Therefore, it is expected that the analytical estimate of the rate of energy release would also be in agreement with a simulation with much higher resolution.
Therefore, the energy released during photospheric cancellation can be estimated accurately with knowledge of the parameters of the cancellation.

We also studied the atmospheric response to the reconnection and found that both cool and hot ejections may be generated. In this work we have mainly focused on how the formation of the different ejections depends on the height of the energy release. Furthermore, we  comment briefly on the acceleration of the ejections, without  going into too much detail as this is beyond the scope of the current paper. In the vicinity of the reconnection region, shocks accelerate the reconnection outflows. 
These outflows collide with the overlying horizontal ambient field which diverts them sideways. 
There, further shocks may develop due to an increase in the local compression. The diverted flows are the main ejections discussed in this paper.

In addition, the collision of the reconnection outflow with the overlying horizontal ambient field accelerates the plasma of the overlying field.
This process is more important when the reconnection occurs lower in the atmosphere, as it accelerates denser material, forming another cool ejection.
When the reconnection occurs lower in the atmosphere both the diverted reconnection outflow and the displaced dense material can be cool, but the density and temperature of the two can be quite different. Similar results were obtained in the 2D simulations of Part II. The acceleration of the ejections is similar to what is found in 2.5D numerical simulations 
of reconnection at different atmospheric heights driven by flux emergence  \citep{Takasao_etal2013}.

The temporal evolution and  plasma properties of the jets were also examined, validating our previous results and extending them to include three-dimensional spatial effects. In summary, our results indicate that, depending on the properties of the cancelling region: i) either hot ejections or cool ones or a combination of both hot and cool ejections can be formed, ii) these can be formed with or without a time difference and iii) with or without a spatial offset.
The spatial offset is simply due to the fact that the nulls along the ring are located at different heights. 
The time difference can occur both due to reconnection occurring at different spatial locations, where a different amount of energy is needed to energise plasma of different densities leading to a time difference between the ejections, and due to current sheet moving lower in the atmosphere later in the cancellation. In addition, note that the current sheet has a vertical length, and therefore reconnection occurs at multiple heights at any given time along the vertical extend of it.
We note that different kinds of ejection can be formed (e.g., either hot or cool or both) when the reconnection at different atmospheric heights is driven by flux emergence  \citep[e.g.,][]{Shibata_1999}.

The hot ejections in our numerical investigation have temperatures of $0.6-2$~MK, densities of 10$^{-13} - 10^{-15}$ g cm$^{-3}$, velocities of $25-125$~km s$^{-1}$ and widths  $4-8$ Mm. The hot ejections always reach the boundary of the domain, and so their lengths are at least $12$ Mm. Such values are typical for small-scale coronal and transition region jets. 
The cool ejections have temperatures of $0.01-0.2$ MK, densities of 10$^{-12} - 10^{-14}$ g cm$^{-3}$, velocities of $5-50$~km s$^{-1}$, widths of $2-4$ Mm and lengths of $3-10$ Mm. These are typical values for surges. In addition, in case 5 we find that an even cooler ejection of photospheric or chromospheric temperature (around 6300 K) is produced, as described in the previous paragraph.
In some of the cases, ejections of intermediate temperatures of $0.2-0.6$~MK are found as the null progressively moves towards the photosphere. 
The above values are similar to the ones reported for hot and cool ejections associated with surges and hot jets driven by reconnection during flux emergence \citep[see e.g.,][]{morenoinsertis13,MacTaggart_etal2015,nobrega_etal2016}. 
We expect that plasmoid-induced reconnection \cite[e.g.,][]{Peter_etal2019} can affect the temperature and density of the jets, but our main conclusion about the time differences and spatial offsets of the outflows should not be affected.
The physical size and properties of these jets could also be affected by the addition of an oblique ambient field and by changing the properties of the cancelling polarities.
For example, the width of ejections will depend on the extent of the current sheet along which reconnection occurs, which will depend on the sizes and fluxes of the photospheric polarities, in addition to the ambient field strength and direction. 

Note furthermore that, in cases where the photospheric polarities are driven for a smaller time, mimicking a shorter-lasting flux cancellation, the null point would reconnect for a shorter time. Therefore, depending on the properties of the reconnection region, these outflows could be shorter-lasting and associated with a shorter burst of energy release.
In principle, depending on the specifics of the cancelling region (e.g., the local plasma $\beta$ distribution, the photospheric fluxes, the sizes of the polarities, the convergence speed, the flux cancellation rate, ambient magnetic field strength and orientation, etc), various combinations of the manifestations of the ejections found in our simulations could occur.

So, our results demonstrate that, in cancelling regions, a wide spectrum of ejections can be produced, with a wide range of different temperatures and sizes associated with Ellerman bombs, UV bursts and IRIS bombs and other hot and cool ejections  \citep{Yang_etal2013,Vissers_etal2015,Reid_etal2016,Rutten_2016,Nelson_etal2017,Rouppe_van_der_Voort2017,Young_etal2018,Tian_etal2018,Chen_etal2019,Guglielmino_etal2019,Ortiz_etal2020,Tiwari_etal2019,Huang_etal2019,Madjarska_etal2019}. 
The predicted spatial offsets and time delays for hot and cool ejections are in accordance with observational findings. E.g., surges and EUV/X-ray jets with time delays of $\sim 5-10$ min have been recorded \citep{Schmieder_etal1994,Chae_etal1999,Alexander_Fletcher_1999,Jiang_etal2007}. Also, hot and cool ejections can appear to be co-temporal and be either spatially offset or co-spatial \citep[e.g.,][]{Mulay_etal2017,Kontogiannis_etal2020}, or only cool ejections can form\citep[e.g., surges, ][]{Ortiz_etal2020}.
We note further that similar energetic events at the feet of coronal loops could possibly create jet-like structures observed at these locations \citep{chitta17a,chitta17b}. 
In addition, given the increasing evidence for the presence of unresolved minority polarities in the photosphere \cite[e.g.,][]{solanki17b,smitha17}, such ejections resulting from flux cancellation can give rise to jet-like features found inside plages and strong network associated with the unresolved underlying minority polarities \citep{Wang_etal2016, Wang_etal2019}. We expect that other small-scale phenomena, such as spicules, could be associated with  reconnection at lower atmospheric heights between converging cancelling polarities, and can potentially be associated with the cool outflows predicted from our model for different values of  parameter space \citep[e.g.,][]{Samanta_etal2019}.

Our results show that the local plasma $\beta$ is important for the production of outflows and energy bursts. Because reconnection is driven along the part of the ring of nulls located in a low-$\beta$ plasma region, only that region will produce jets, and thus, the spatial extent of the jets will depend on the local plasma $\beta$. 
A detailed assessment of the role of the local plasma $\beta$ on the spatial extent of the jets producing by a cancelling region would require a different modelling approach than the one employed here.
We have imposed an initial  potential field  in a simple stratified atmosphere, for which the plasma density and temperature in the magnetised region simply follows the stratification. 
However, self-consistent flux emergence simulations show that the plasma density and temperature in an emerged field can be very different from the background stratification, since the adiabatic expansion of the field naturally forms cooler ``bubbles'' of magnetised plasma \cite[e.g.][]{Archontis_etal2004, Leake_etal2013b,Nobrega-siverio_etal2020}.
During our numerical experiment, we have not incorporated such an effect.
Simulations where the polarities emerge self-consistently, interact across an extended vertical current sheet and produce transient jet-like events \citep[e.g.][]{Archontis_etal2013,Syntelis_etal2015}, Ellerman bombs \cite[e.g.][]{Archontis_etal2009b, Danilovic_etal2017}, UV bursts, and the spectral imprints of the resulting hot and cool outflows \citep{Hansteen_etal2017,Hansteen_etal2019}, are better suited for a detailed study of the role of plasma $\beta$ on the spatial extent of the jets.

In this work, we have further validated our analytical theory using 3D numerical computations, providing support to our suggestion that nanoflares driven by magnetic flux cancellation can be an important mechanism for chromospheric and coronal heating as proposed in Paper I, built upon the recent observational findings.
Furthermore, our models suggest that a plethora of different hot and cool outflows produced with/without time difference and spatial offset in small-scale cancelling regions, represent a wide range of different manifestations of cancellation-driven reconnection.
In future, we aim to extend our model by including different orientations of the horizontal field, including oblique fields, and examining different magnetic configurations.

\acknowledgements

This research has made use of NASA's Astrophysics Data System.  
This work was supported by computational time granted from the Greek Research and Technology Network (GRNET) in the National HPC facility – ARIS. PS acknowledge support by the ERC synergy grant ``The Whole Sun''.

\bibliographystyle{aasjournal} 
\bibliography{bibliography}

\end{document}